\documentclass[aps,amsmath,notitlepage,twocolumn,amssymb,prl,longbibliography]{revtex4-1}

\usepackage{graphicx}
\usepackage{dcolumn}
\usepackage{bm}
\usepackage[colorlinks=true,citecolor=red,urlcolor=blue]{hyperref}
\usepackage{wasysym}
\usepackage{stmaryrd}
\usepackage{verbatim}
\usepackage{subfigure}
\usepackage{amsmath}

\begin{document}
\title{Asymmetric Ferromagnetic Criticality in Pyrochlore Ferromagnet {Lu$_2$V$_2$O$_7$}}

\author{N. Su$^{1,6}$}
\thanks{These authors contributed equally to this work.}
\author{F. -Y. Li$^{2}$}
\thanks{These authors contributed equally to this work.}
\author{Y. Y. Jiao$^{1,6}$}
\thanks{Present address: Faculty of Science, Wuhan University of Science and Technology, Wuhan, Hubei 430062, China}
\author{Z. Y. Liu$^{1,3}$}
\author{J. P. Sun$^{1,6}$}
\author{B. S. Wang$^{1,6,7}$}
\author{Y. Sui$^{3}$}
\author{H. D. Zhou$^{4}$}
\author{G. Chen$^{2,5}$}
\email{gangchen.physics@gmail.com}
\author{J. -G. Cheng$^{1,6,7}$}
\email{jgcheng@iphy.ac.cn}
\affiliation{$^{1}$Beijing National Laboratory for Condensed Matter Physics and Institute of Physics, Chinese Academy of Sciences, Beijing 100190, China\\
$^{2}$State Key Laboratory of Surface Physics and Department of Physics, 
Fudan University, Shanghai 200433, China\\
$^{3}$Department of Physics, Harbin Institute of Technology, Harbin 150001, China\\
$^{4}$Department of Physics and Astronomy, University of Tennessee, Knoxville, TN 37996, USA\\
$^{5}$Department of Physics and Center of Theoretical and Computational Physics,
The University of Hong Kong, Pokfulam Road, Hong Kong, China\\
$^{6}$School of Physical Sciences, University of Chinese Academy of Sciences, Beijing 100190, China\\
$^{7}$Songshan Lake Materials Laboratory, Dongguan, Guangdong 523808, China}

\date{\today}

\begin{abstract}
Critical phenomenon at the phase transition reveals the universal and 
long-distance properties of the criticality. 
We study the ferromagnetic criticality of the pyrochlore magnet 
{Lu$_2$V$_2$O$_7$} at the ferromagnetic transition ${T_\text{c}\approx 70\, \text{K}}$
from the isotherms of magnetization $M(H)$ via an iteration 
process and the Kouvel-Fisher method. The critical exponents 
associated with the transition are determined as 
${\beta = 0.32(1)}$, ${\gamma = 1.41(1)}$, and ${\delta = 5.38}$. 
The validity of these critical exponents is further verified by 
scaling all the $M(H)$ data in the vicinity of $T_\text{c}$ onto 
two universal curves in the plot of $M/|\varepsilon|^\beta$ versus $H/|\varepsilon|^{\beta+\gamma}$, 
where ${\varepsilon = T/T_\text{c} -1}$. The obtained $\beta$ and $\gamma$ values show 
asymmetric behaviors on the ${T < T_\text{c}}$ and the ${T > T_\text{c}}$ sides, 
and are consistent with the predicted values of 3D Ising and cubic universality classes, 
respectively. This makes {Lu$_2$V$_2$O$_7$} a rare example in which the critical 
behaviors associated with a ferromagnetic transition belong to different universality classes. 
We describe the observed criticality from the Ginzburg-Landau theory with 
the quartic cubic anisotropy that microscopically originates from the anti-symmetric 
Dzyaloshinskii-Moriya interaction as revealed by recent magnon thermal 
Hall effect and theoretical investigations.
\end{abstract}

\maketitle

As a representative subject of quantum magnetism, pyrochlore antiferromagnets have 
attracted a significant attention in recent years~\cite{RevModPhys.82.53}. 
Many interesting phenomena including classical spin ice~\cite{Castelnovo12008,Gingras2001}, 
quantum spin ice~\cite{PhysRevLett.98.157204,Gingras2014,Ross11}, 
pyrochlore ice U(1) spin liquid~\cite{Hermele04,Savary12,Gingras2014,Sungbin2012,Huang2014}, 
quantum order by disorder~\cite{PhysRevLett.109.167201,conmatphys}, 
spin nematics~\cite{PhysRevX.7.041057,PhysRevB.81.184409}, 
symmetry enriched topological orders~\cite{GangChen2017,PhysRevB.96.195127,Huang2014}, 
topological magnetic excitations~\cite{Weylmagnon,PhysRevB.98.045109,PhysRevB.97.115162} 
have been proposed and/or discovered for various 
compounds in the pyrochlore antiferromagnet families. 
While most efforts of this field have been devoted to the antiferromagnets, 
the pyrochlore ferromagnet {Lu$_2$V$_2$O$_7$} may stand out  
in the field of pyrochlore magnets by providing some rather 
unique and robust phenomena~\cite{ref1,ref2,ref3,ref4,ref6,ref7}. 
{Lu$_2$V$_2$O$_7$} is a vanadium based pyrochlore Mott insulator  
that orders ferromagnetically below about 70 K. Although
it is a conventional ferromagnet, this material shows a
remarkable magnon thermal Hall transport under magnetic
field. Microscopically, the large thermal Hall effect~\cite{ref6,ref7} in this material
is attributed to the non-trivial Berry curvature of the thermally populated magnon 
bands that originate from the antisymmetric Dzyaloshinskii-Moriya (DM) 
interaction~\cite{ref6,ref7,ref9,ref10} between the V$^{4+}$ 
spin-1/2 localized moments on the pyrochlore lattice. Since this discovery, 
{Lu$_2$V$_2$O$_7$} has become one of the stereotypes for the thermal 
Hall transports in Mott insulating materials.  

To motivate our work, we here provide a general discussion about 
the magnetic properties of the pyrochlore ferromagnet {Lu$_2$V$_2$O$_7$} 
or any conventionally ordered matter. For the conventional 
ordered states, there are several basic and phenomenological aspects. 
The first one would be the order parameter and the ordering structure. 
{Lu$_2$V$_2$O$_7$} is a conventional ferromagnet.  
Being a ferromagnet is a global static property, the system should have a 
non-collinear spin structure within the four-sublattice unit cell. 
This is a direct consequence of the DM interaction. 
The second one would be the dynamic property or fluctuation with respect
to the magnetic order. The thermal Hall transport is an overall effect 
caused by the magnetic excitations and reflects the weighted average of 
the Berry curvatures over the magnon bands~\cite{ref6,ref7}. 
More detailed information 
about the dynamics would come from the magnon dispersion directly.  
This could be obtained through inelastic neutron scattering measurements. 
Theoretically, it has been suggested that the Weyl magnons~\cite{Weylmagnon} may be
present from studying a minimal model for {Lu$_2$V$_2$O$_7$} and 
can be one origin for the thermal Hall effect in this system~\cite{ref8,PhysRevLett.117.157204}. 
The third aspect is the critical property due to the flucutations 
of order parameters associated with this ferromagnetic transition 
in {Lu$_2$V$_2$O$_7$}. This aspect has not been carefully considered. 
Given the growing interests in this simple ferromagnetic pyrochlore oxide, 
we are motivated to explore its critical behaviors around $T_\text{c}$ 
via analyzing the isotherms of magnetization $M(H)$ with an iteration process 
and the Kouvel-Fisher method. Remarkably, we find that the critical exponents 
on both sides of $T_{\text c}$ are not equal and show 3D Ising-like and 
3D cubic-like universality classes on each side. 
This differs from the conventional wisdom about the criticality that 
the asymmetry usually occurs in the non-universal prefactors of the scaling 
law rather than the scaling exponents. We discuss this asymmetric
ferromagnetic criticality within Ginzburg-Landau theory and provide
our view on the microscopic properties of the V-based pyrochlore magnet 
{Lu$_2$V$_2$O$_7$}. We further give a general comment about the spin-orbit 
coupling and the Kitaev physics in the V-based magnets. 

The critical behaviors of {Lu$_2$V$_2$O$_7$} around the ferromagnetic transition 
can be described by a series of critical exponents $\beta$, $\gamma$, and $\delta$ 
that reflect the effective magnetic interactions at play~\cite{ref13}. Different critical 
exponents have been derived theoretically for different models,
\textit{e.g.} ${\beta = 0.365}$ and ${\gamma = 1.386}$ for 3D Heisenberg model, 
$\beta = 0.345$ and $\gamma = 1.316$ for 3D XY model, and $\beta = 0.325$ and 
$\gamma = 1.24$ for 3D Ising model, respectively~\cite{Kaul}. 
These exponents are obtained by analyzing the isothermal magnetizations $M(H)$ 
near $T_\text{c}$, \textit{viz}. 
\begin{align}
	M_\text{s}(T) &\propto (T_\text{c}-T)^\beta &\text{for \quad} T < T_\text{c},  
	\label{eq1} \\
	\chi_0^{-1}(T) &\propto (T-T_\text{c})^\gamma &\text{for \quad} T > T_\text{c},
	\label{eq2} \\
	M(H) &\propto H^{1/\delta} &\text{for \quad} T = T_\text{c},
	\label{eq3}
\end{align}
where $M_\text{s}$ is the spontaneous magnetization and $\chi_0^{-1}$ is the inverse initial magnetic susceptibility, respectively. In a previous work, Zhou \textit{et al.}~\cite{ref14} have obtained 
the critical exponents ${\beta = 0.42}$ and ${\gamma = 1.85}$ for {Lu$_2$V$_2$O$_7$} and ascribed 
them to a 3D Heisenberg model. However, these values show a large deviation from those predicted 
theoretically. In addition, the straight lines in the modified Arrott plot, $M^{1/\beta}$ versus 
$(H/M)^{1/\gamma}$, are not parallel at all. Here, we reinvestigate its critical behaviors around 
$T_\text{c}$ via analyzing the isotherms of magnetization $M(H)$ with an iteration process and the 
Kouvel-Fisher method. Interestingly, we found that the obtained ${\beta = 0.322}$ at ${T < T_\text{c}}$ 
and ${\gamma = 1.41}$ at $T > T_\text{c}$ are close to the predicted values of 3D Ising and 
3D Heisenberg or cubic universality classes, respectively. 
{Lu$_2$V$_2$O$_7$} single crystals used in the present study were grown with the traveling-solvent 
floating-zone technique. Details about the crystal growth and sample characterizations have been 
published elsewhere~\cite{ref14}. The magnetic properties of {Lu$_2$V$_2$O$_7$} were measured with 
a commercial Magnetic Property Measurement System (MPMS-III, Quantum Design).  

\begin{figure}[t]
	\includegraphics[width=8.5cm]{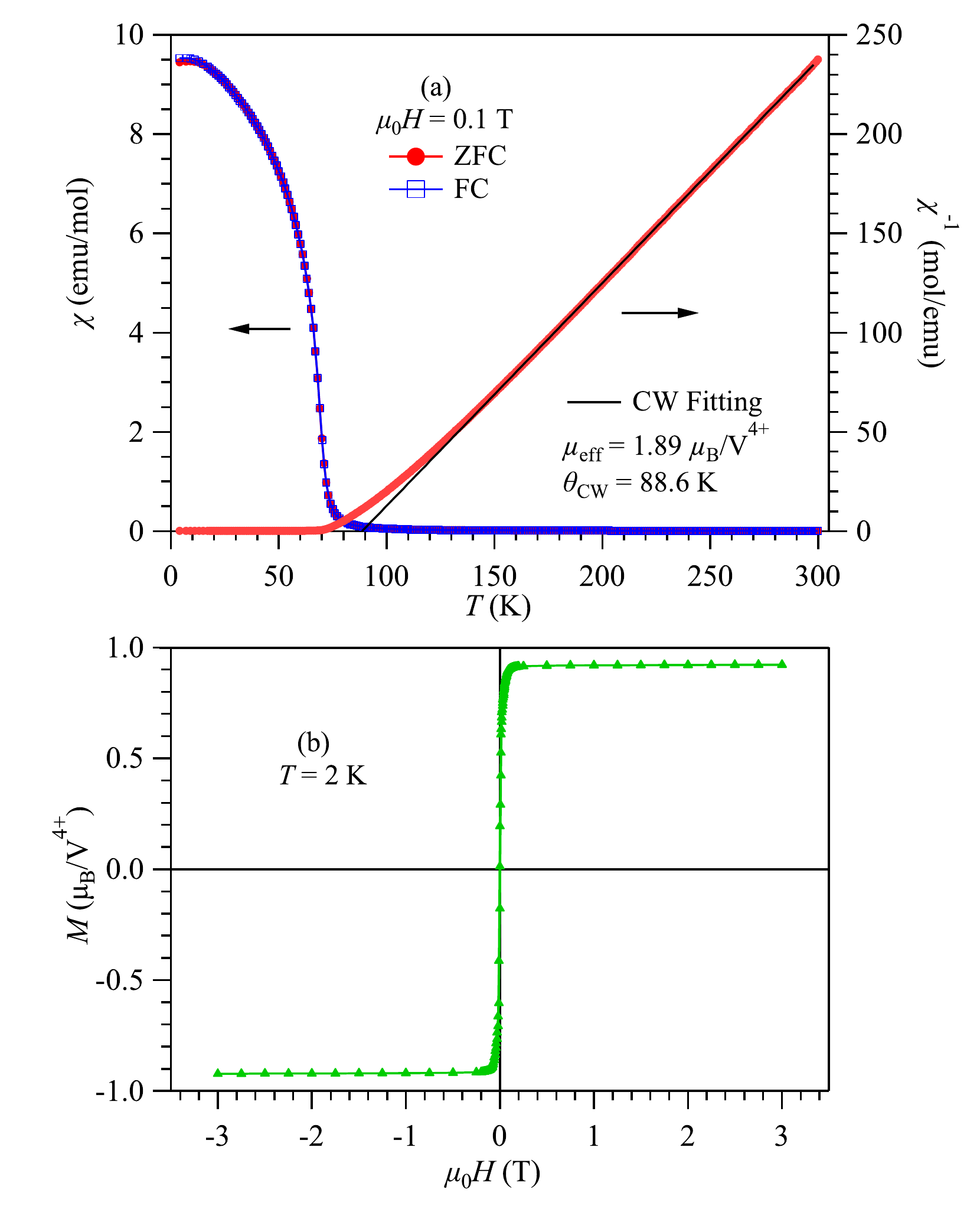}
	\caption{(Color online.) (a) Temperature dependence of dc magnetic susceptibility $\chi(T)$ and its inverse $\chi^{-1}(T)$ measured in both zero-field-cooled (ZFC) and field-cooled (FC) modes under $\mu_0H = 0.1$ T. The solid line is the Curie-Weiss (CW) fitting curve. (b) Isothermal magnetization $M(H)$ curve at 2 K. 
	} 
	\label{fig1}
\end{figure}

In Fig.~\ref{fig1}(a), we plot the temperature dependence of dc magnetic susceptibility $\chi(T)$ 
and its inverse $\chi^{-1}(T)$ measured under ${\mu_0 H = 0.1}$ T in both zero-field-cooled 
(ZFC) and field-cooled (FC) modes. As can be seen, the ZFC and FC $\chi(T)$ curves are 
overlapped with each other and the ferromagnetic transition around ${T_\text{c} \approx 70}$ K 
can be clearly visible from the sharp rise of $\chi(T)$. In the paramagnetic region above 
$T_\text{c}$, we have applied the Curie-Weiss (CW) fitting to $\chi^{-1}(T)$ in 
the temperature range 150-300 K and extracted the effective moment of 
${\mu_\text{eff} = 1.89} \,{\mu_\text{B}/\text{V}^{4+}}$ 
and the Weiss temperature ${\theta_\text{CW}= 88.6}$ K. The obtained $\mu_\text{eff}$ 
is closed to expected value of $1.73 \,\mu_\text{B}$ for ${S = 1/2}$ of V$^{4+}$, 
and the deviation from this ideal value originates from the spin-orbit coupling 
of the V$^{4+}$ ion. The positive $\theta_\text{CW}$ signals the dominant ferromagnetic 
exchange interactions in this system. Fig.~\ref{fig1}(b) displays the $M(H)$ curve at 2 K, 
which exhibits a typical ferromagnetic behavior and reaches a saturation moment 
of ${\sim1.0 \,\mu_\text{B}}$ as expected. All these results are consistent with 
those reported previously and confirm the high quality of the studied crystal~\cite{ref14}.

\begin{figure}[t]
	\includegraphics[width=8.5cm]{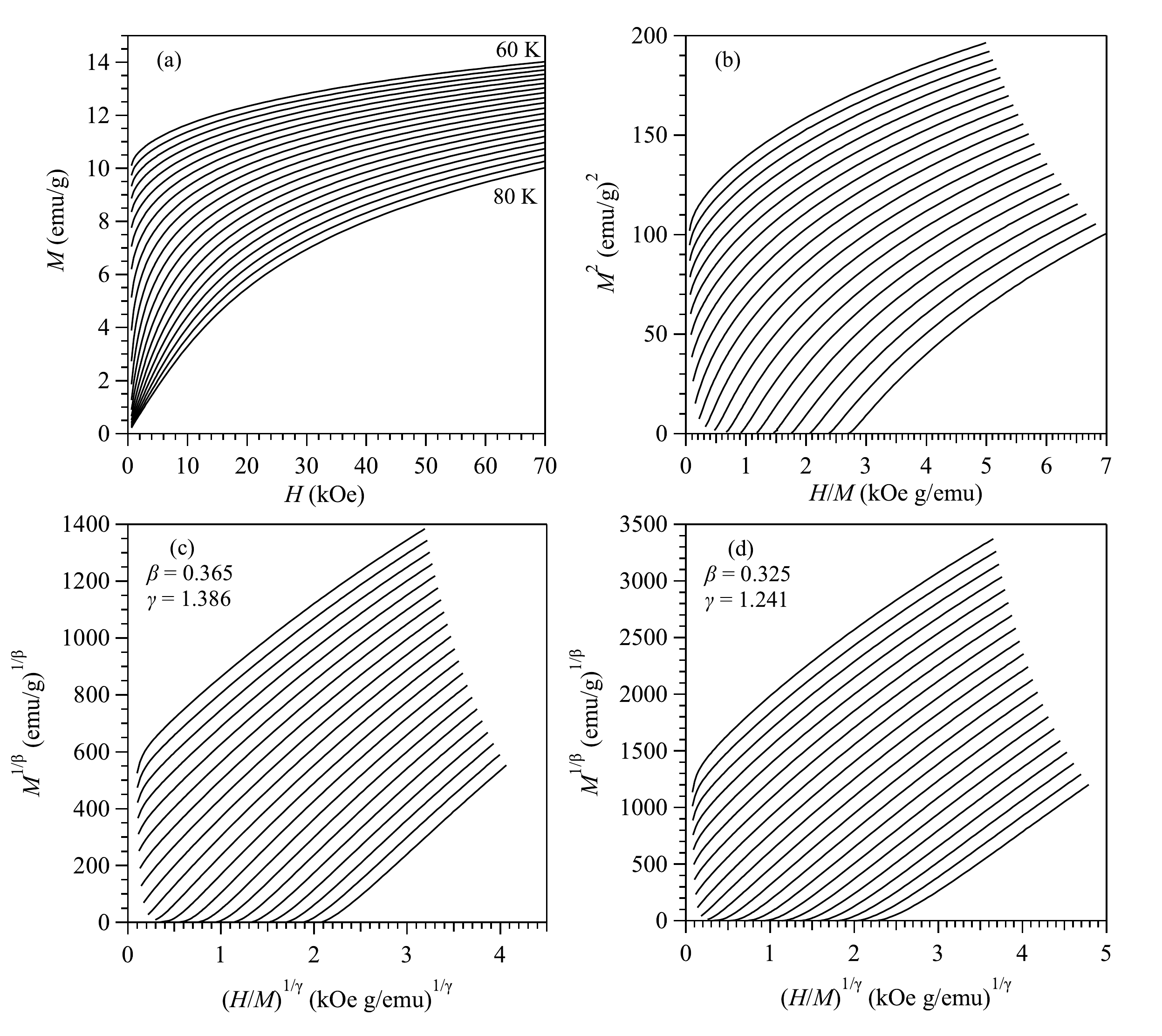}
	\caption{(Color online.) (a) Isothermal magnetization curves between 60 and 80 K, and the modified Arrott plots of these curves with critical exponents of (b) mean field model $\beta$ = 0.5, $\gamma$ = 1, (c) 3D Heisenberg model $\beta$ = 0.365, $\gamma$ = 1.386, (d) 3D Ising model $\beta$ = 0.325, $\gamma$ = 1.241.
	} 
	\label{fig2}
\end{figure}

In Fig.~\ref{fig2}(a), we plot the isothermal $M(H)$ curves of {Lu$_2$V$_2$O$_7$} 
in the temperature range of 60-80 K, which covers the ferromagnetic transition. 
The demagnetization effect has been corrected. These $M(H)$ data are replotted 
in the Arrott plot $M^2$ vs $H/M$ in Fig.~\ref{fig2}(b), and in the modified 
Arrott plots $M^{1/\beta}$ vs $(H/M)^{1/\gamma}$ with the critical exponents 
of 3D Heisenberg and 3D Ising models in Fig.~\ref{fig2}(c) and (d), respectively. 
The curved Arrott plot in Fig.~\ref{fig2}(b) rules out the possibility of mean-field model, 
but the positive slope of the $M^2$ vs $H/M$ confirms that the paramagnet-ferromagnet 
transition is a continuous transition. On the other hand, the modified Arrott plots in 
Fig.~\ref{fig2}(c) and (d) gave roughly parallel straight lines, 
and it is hard to distinguish visually which model could better 
describe the ferromagnetism of {Lu$_2$V$_2$O$_7$}.

\begin{figure}[t]
\includegraphics[width=8cm]{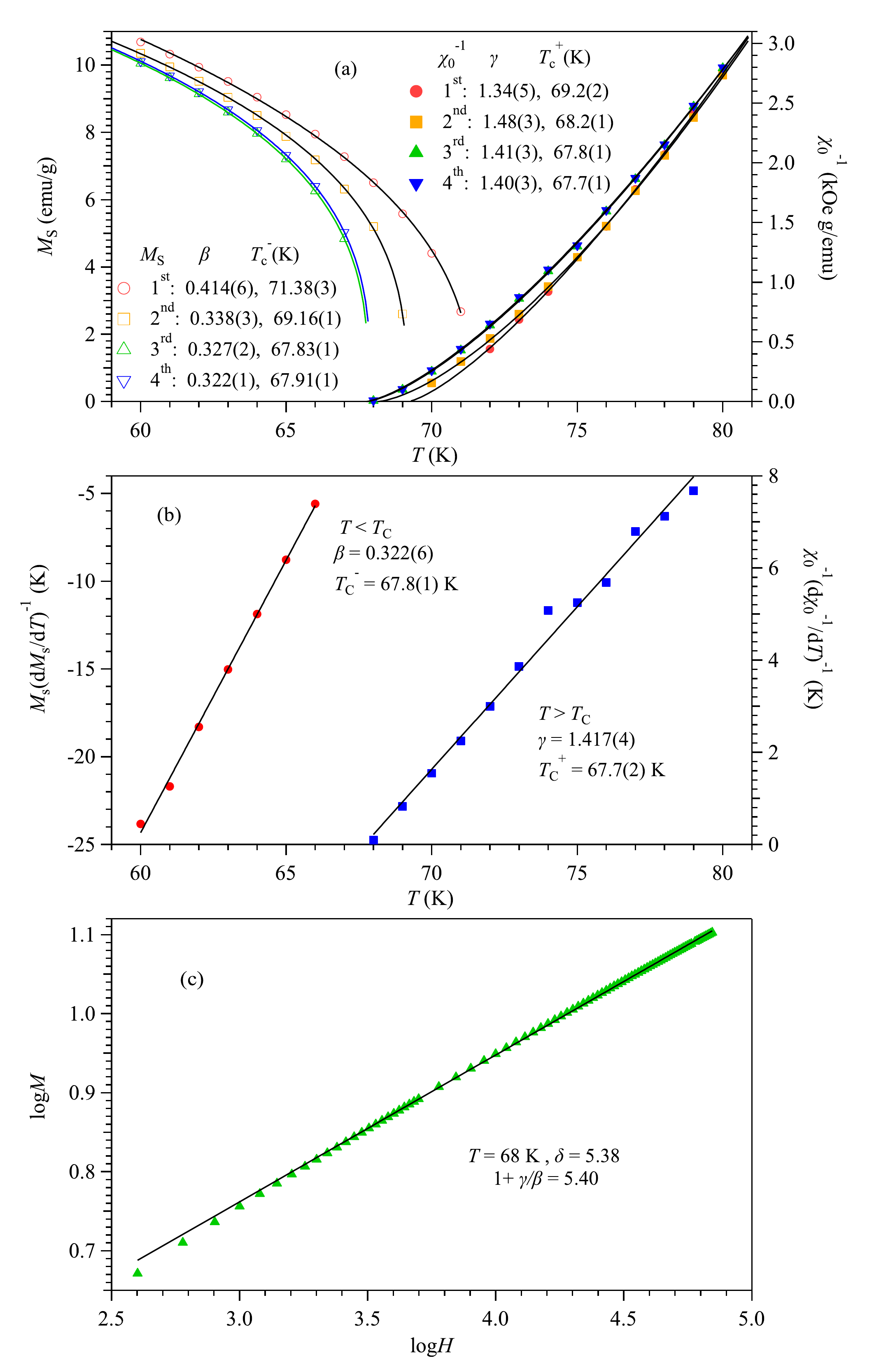}
\caption{(Color online.) Critical exponents $\beta$ and $\gamma$, 
and critical temperatures $T_\text{c}^-$ and $T_\text{c}^+$ determined 
from (a) an iteration process started from the mean-field Arrott plot, 
and (b) Kouvel-Fisher plots. (c) Critical isotherm at $T$ = 68 K in a double 
logarithmic plot and a linear fitting to extract the critical exponent $\delta$. 
The Widom scaling relation, $\delta=1+\gamma/\beta$.
} 
\label{fig3}
\end{figure}

In order to determine precisely the critical exponents, we employed an iteration process in analyzing the isothermal $M(H)$ data near $T_\text{c}$ based on the general formula \eqref{eq1} to \eqref{eq3} given above~\cite{ref15, ref16}. Starting from the Arrott plot shown in Fig.~\ref{fig2}(b), we obtain the first set of $M_\text{s}(T)$ and $\chi_0^{-1}(T)$ by extrapolating the corresponding $M^2$ vs $H/M$ curves to the vertical and horizontal axes, respectively. As shown in Fig.~\ref{fig3}(a), the obtained $M_\text{s}(T)$ and $\chi_0^{-1}(T)$ are fitted with the Eqs.~\eqref{eq1} and \eqref{eq2}, respectively, to extract the first set of critical exponents and critical temperatures, i.e. ${\beta = 0.414(6)}$, ${T_\text{c}^- = 71.38(3)}$ K, 
and ${\gamma = 1.34(5)}$, ${T_\text{c}^+ = 69.2(3)}$ K as listed in the figure. 
By using the obtained $\beta$ and $\gamma$ values, we then construct a modified 
Arrott plot $M^{1/\beta}$ vs $(H/M)^{1/\gamma}$ and repeat the above process to 
obtain the second set of $M_\text{s}(T)$ and $\chi_0^{-1}(T)$ and the corresponding 
critical exponents and critical temperatures. As illustrated in Fig.~\ref{fig3}(a), 
after three iterations the fitting parameters are converged to ${\beta = 0.322(1)}$, 
${T_\text{c}^- = 67.91(1)}$ K, and ${\gamma = 1.402(3)}$, ${T_\text{c}^+ = 67.7(1)}$ K. 
These critical exponents are different from those reported previously~\cite{ref14}.

We further determine the critical exponents by employing the Kouvel-Fisher 
relation~\cite{ref17}, \textit{viz}.
\begin{align}
M_\text{s}(T)[dM_\text{s}/dT]^{-1}&=\frac{T-T_\text{c}^-}{\beta}, \label{eq4}
\\
\chi_0^{-1}(T)[d\chi_0^{-1}/dT]^{-1}&=\frac{T-T_\text{c}^-}{\gamma}, \label{eq5}
\end{align}
in which the $M_\text{s}(T)$ and $\chi_0^{-1}(T)$ were obtained from the modified 
Arrott plot with the final critical exponents obtained above. As shown in 
Fig.~\ref{fig3}(b), linear fittings to the plots of $M_\text{s}[dM_\text{s}/dT]^{-1}$ 
and $\chi_0^{-1}[d\chi_0^{-1}/dT]^{-1}$ versus $T$ yield ${\beta = 0.322(6)}$, 
${T_\text{c}^- = 67.8(1)}$ K, and ${\gamma = 1.417(4)}$, ${T_\text{c}^+ = 67.7(2)}$ K. 
Both values of $\beta$ and $\gamma$ obtained by the Kouvel-Fisher relation agree 
well with results from the iterations of modified Arrott plot, confirming the 
validity of these above analysis. For the completeness, we also estimate the 
critical exponent $\delta$. In Fig.~\ref{fig3}(c), we display the double 
logarithmic plot of $M$ vs $H$ at 68 K, which is very close to the critical 
temperature ${T_\text{c} = 67.8(1)}$ K determined above. As can be seen, 
the data falls nearly on a straight line with a slope of $1/\delta$ 
according to Eq.~\eqref{eq3}, and a linear fitting to the data at 
high-field region gives ${\delta = 5.38}$, which fulfills the Widom 
scaling relation perfectly~\cite{ref18}, \textit{i.e.} 
${\delta=1+\gamma/\beta}$ by using ${\gamma = 1.417}$ 
and ${\beta = 0.322}$ obtained above.
 
\begin{figure}[t]
\includegraphics[width=8cm]{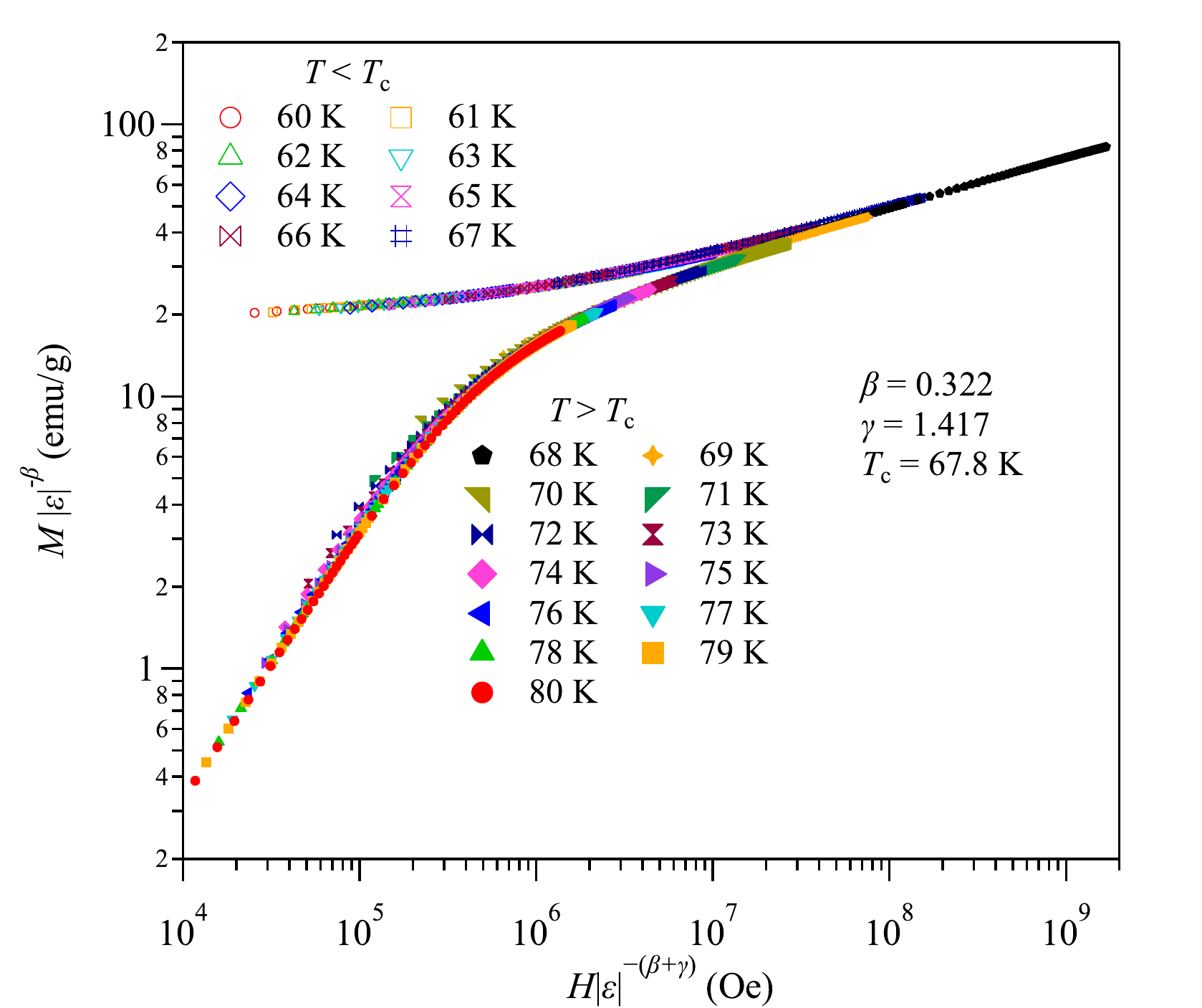}
\caption{(Color online.) Scaling plot for {Lu$_2$V$_2$O$_7$} below and 
above $T_\text{c}$ based on the critical temperature $T_\text{c}$ = 67.8 K 
and $\beta$ = 0.322 and $\gamma$ =1.417.
	} 
	\label{fig4}
\end{figure}

To check the reliability of our analysis on the critical behavior of {Lu$_2$V$_2$O$_7$}, 
we test the obtained critical exponents according to the prediction of the scaling 
hypothesis~\cite{ref13}. In the critical asymptotic region, 
the magnetic equation of state could be expressed as
\begin{equation}
	M (H, \varepsilon) = |\varepsilon|^\beta f_\pm (H/|\varepsilon|^{\beta+\gamma}),       \label{eq6}
\end{equation}
where $f_+$ for ${T > T_\text{c}}$ and $f_-$ for ${T < T_\text{c}}$ 
are regular analytical functions, and ${\varepsilon = T/T_\text{c}-1}$ 
is the reduced temperature. Eq.~\eqref{eq6} implies that for the right 
choice of $\beta$, $\gamma$, and $\delta$ values, $M/|\varepsilon|^\beta$ 
as a function of $H/|\varepsilon|^{\beta+\gamma}$ should produce two universal 
curves: one for ${T > T_\text{c}}$ and the other for ${T < T_\text{c}}$. 
By using the values of $\beta$ and $\gamma$ obtained by the Kouvel-Fisher method and 
${T_\text{c} = 67.8}$ K, we have obtained the scaling plot shown in
Fig.~\ref{fig4}. It is clearly seen that all the points indeed collapse 
into two separate branches. These well-scaled curves in Fig.~\ref{fig4} 
thus further confirm the reliability of the obtained critical exponents.

We have listed the critical exponents of {Lu$_2$V$_2$O$_7$} in Table~\ref{tab1} 
and compared with those of ferromagnetic perovskite {YTiO$_3$}~\cite{ref19} as 
well as the theoretical values from different models~\cite{Kaul, PhysRevB.61.15136}. 
Firstly, we found that the critical exponents for these two ferromagnetic 
oxides with ${S = 1/2}$ are stunningly similar with each other. Secondly,
 it is interesting to note that their $\beta$ and $\gamma$ values are not 
 consistent with those predicted by a single model. Instead, $\beta$ is 
 very close to the predicted value of 3D Ising universality class, 
whereas $\gamma$ is close to the predicated value of 3D Heisenberg or cubic universality class. 
This makes {Lu$_2$V$_2$O$_7$} and {YTiO$_3$} rather rare cases in which 
the critical behaviors above and below $T_\text{c}$ are described by two 
different universality classes. This finding is intriguing because the 
critical behavior associated with a continuous phase transition should 
be described by one universality class with a single set of critical exponents,
and the asymmetric behaviors usually occur in the non-universal prefactors 
of the scaling relations. The transition from the 3D Heisenberg$/$cubic to Ising universality 
class across $T_\text{c}$ implies the reduction of spin dimensionality or 
effective spin components upon the ferromagnetic ordering. 
As mentioned above, recent experimental and theoretical investigations 
on {Lu$_2$V$_2$O$_7$} have revealed the presence of significant anisotropic 
interactions in the form of DM interaction~\cite{ref6,ref7,ref9,ref10}, 
which should be responsible for such a transition. 

\begin{table}[h]
	\begin{tabular}{c|c|c|c|c}		
		\hline\hline
		& \hspace{0cm} $\beta$ \hspace{0cm} & \hspace{0cm} $\gamma$ \hspace{0cm} & \hspace{0cm} $\delta$ \hspace{0cm} & Ref.
		\\ \hline
		{Lu$_2$V$_2$O$_7$} &	0.322 &  1.402	& 5.38	& This work
		\\ \hline
		{YTiO$_3$} & 0.328	& 1.441	& 5.39 & ~\cite{ref19}	
		\\ \hline
		3D Ising  & 0.325	& 1.241	& 4.82 & ~\cite{Kaul}	
		\\ \hline
		3D XY 	& 0.345	& 1.316	& 4.81 & ~\cite{Kaul}
		\\ \hline
		3D Heisenberg 	& 0.365	& 1.386	& 4.8 & ~\cite{Kaul}	
		\\ \hline
		3D Cubic  & 0.364 & 1.390 & 4.82 & ~\cite{PhysRevB.61.15136}
		\\ \hline\hline
	\end{tabular}
	\caption{Critical exponents of the single-valent $S = 1/2$ ferromagnetic oxides, {Lu$_2$V$_2$O$_7$} with the pyrochlore structure and {YTiO$_3$} with the perovskite structure, in comparison with the theoretical values from different models.}
	\label{tab1}
\end{table}

The observation of similar critical behaviors in {Lu$_2$V$_2$O$_7$} 
and {YTiO$_3$} is not completely unexpected since both V$^{4+}$ and Ti$^{3+}$ ions 
have identical 3d$^1$ electronic configuration with an active orbital degree of 
freedom in the octahedrally coordinated environments. The active orbitals are
the lower $t_{2g}$ orbitals for both compounds. For {Lu$_2$V$_2$O$_7$}, we know 
that the trigonal distortion would further split the $t_{2g}$ orbitals into $a_{1g}$ 
orbital and $e_{2g}'$ orbitals. On the other hand, the atomic spin-orbit coupling 
is active for the $t_{2g}$ orbitals. The spin-1/2 local moment of the V$^{4+}$ ion
should be interpreted as the effective spin-1/2 of the ground state Kramers doublet
for a local single-ion Hamiltonian with the spin-orbit coupling and the trigonal distortion. 
From this perspective, the nature of the V$^{4+}$ local moment would be similar to the 
one for the Ir$^{4+}$ ion, where the latter can be thought as a 5d$^1$ hole~\cite{PhysRevB.78.094403} 
(instead of 3d$^1$ electron for V$^{4+}$). An interesting consequence of this correspondence 
is that, the rich physics due to the spin-orbit entanglement for iridates, such as
Kitaev interaction~\cite{PhysRevLett.102.017205} and/or highly anisotropic exchange~\cite{PhysRevB.78.094403}, 
may be found among the 3d transition metal oxides like vanadates. 

The orbital nature of the V$^{4+}$ local moment is consistent with the presence  
of significant (anti-symmetric) DM interaction. Moreover, 
for the same reason, we expect that the symmetric pseudo-dipolar interaction 
should be present and equally important as the DM interaction. 
This may not effect certain topological properties of the magnon bands too much 
as these properties are topologically robust. Since the local moment is effective 
spin-1/2, the single-ion anisotropy should not be present. 
As for the universal long-distance
property in the vicinity of the transition, one does not need to worry too much
about the detailed form of the interaction. Due to the orbital character of the local
moment, the lattice symmetry directly acts on the effective spin components that 
effectively reduces the ferromagnetic order parameters from having an O(3) symmetry to the cubic symmetry. From the Ginzburg-Landau symmetry analysis, 
we propose the following action to describe the behavior of 
{Lu$_2$V$_2$O$_7$} near the transition,

\begin{align}
\begin{split}
	\mathcal{L}=&\, a|\boldsymbol{M}|^2+b|\boldsymbol{M}|^4\\
	&+\lambda_\text{c}\big[(M^x)^4+(M^y)^4+(M^z)^4\big]+\cdots,
\end{split}
\end{align}

where $\boldsymbol{M}$ is the coarse-grained ferromagnetic order parameter,
and ``$\cdots$'' represents the higher-order terms neglected here.
The first two terms are isotropic with $a\approx a'(T-T_\text{c})$, $a'>0$ and $b>0$.
They can describe the conventional ferromagnetic transition for $O(3)$-type order parameter,
belong to the 3D Heisenberg universality class.
We include the cubic anisotropy of the {Lu$_2$V$_2$O$_7$} system via 
$\lambda_\text{c}$ and require $\lambda_\text{c}<0$. This cubic anisotropy
then favors the order parameter $\boldsymbol{M}$ to be 
aligned with the $\langle001\rangle$ directions, which is consistent with
the experimental observation~\cite{ref6}. 
For a small $\lambda_\text{c}$, the system should be found 
still in the Heisenberg universality class when $T$ is above
but not too close to $T_\text{c}$.
When $T$ is further tuned to $T_\text{c}$, one expects a 
crossover from the Heisenberg universality class
to the cubic universality class~\cite{PhysRevB.8.4270}. 
We mention that the critical exponents belong to 
these two universality classes are very close, Table~\ref{tab1} 
thus difficult to be distinguished from each other in experiments.
Recent results in literature give ${\gamma=1.3895}$ for 
the Heisenberg class~\cite{guida1998critical} and ${\gamma=1.390}$
for the cubic class~\cite{PhysRevB.61.15136}, 
which are indeed close to the experimental result ${\gamma=1.402}$ 
obtained in this work. At ${T<T_\text{c}}$, the cubic symmetry is 
spontaneously broken. The system has chosen one of the ${\langle001\rangle}$ 
directions and the order parameter becomes Ising-like, 
leading the critical exponent $\beta$ belong to the Ising universality class.

In summary, we have investigated the critical behavior of ferromagnetic pyrochlore {Lu$_2$V$_2$O$_7$} based on the measurements of isothermal magnetization $M(H)$ around $T_\text{c}$. We found that the ${\beta = 0.322}$ 
at ${T < T_\text{c}}$ and ${\gamma = 1.417}$ at ${T > T_\text{c}}$ are close to the predicted values of 3D Ising and cubic universality classes, respectively. This makes {Lu$_2$V$_2$O$_7$} a rare example in which the critical behaviors associated with a ferromagnetic transition belong to different universality classes. We have rationalized the observed criticality from the Ginzburg-Landau theory with the quartic cubic anisotropy that microscopically originates from the anti-symmetric DM interactions.

\emph{Acknowledgments.}---This work is supported by the National Key R\&D Program of China 
(Grant Nos. 2018YFA0305700, 2018YFA0305800, 2016YFA0301001, 2016YFA0300500), the National Natural Science Foundation of China (Grant Nos. 11574377, 11834016, 11874400), the Strategic Priority Research Program and Key Research Program of Frontier Sciences of the Chinese Academy of Sciences (Grant Nos. XDB25000000, XDB07020100 and QYZDB-SSW-SLH013). J.P.S. and Y.Y.J. acknowledge support from the China Postdoctoral Science Foundation and the Postdoctoral Innovative Talent program.

\bibliography{refs}

\end{document}